\documentclass[fleqn,usenatbib]{mnras}

\usepackage[T1]{fontenc}

\DeclareRobustCommand{\VAN}[3]{#2}
\let\VANthebibliography\thebibliography
\def\thebibliography{\DeclareRobustCommand{\VAN}[3]{##3}\VANthebibliography}

\usepackage{graphicx}	        
\usepackage{amsmath}	        
\usepackage{amssymb}	        
\usepackage[dvipsnames]{xcolor} 

\newcommand{\mSun}{\; {\rm M_\odot}}				
\newcommand{\mJup}{\; {\rm M_{Jup}}}				
\newcommand{\mEarth}{\; {\rm M_\oplus}}				
\newcommand{\au}{\; {\rm au}}						
\newcommand{\yr}{\; {\rm yr}}						
\newcommand{\myr}{\; {\rm Myr}}						
\newcommand{\gyr}{\; {\rm Gyr}}						
\newcommand{\percent}{\; {\rm per \; cent}}			
\newcommand{\HD}{{\rm HD} \;}						

\title[Planet-induced gaps in massive debris discs]{Gap carving by a migrating planet embedded in a massive debris disc}

\author[M. F. Friebe et al.]{
Marc F. Friebe\thanks{E-mail: marc.friebe@uni-jena.de},
Tim D. Pearce
and Torsten L\"{o}hne
\\
Astrophysikalisches Institut und Universit\"{a}tssternwarte, Friedrich-Schiller-Universit\"{a}t Jena, Schillerg\"{a}{\ss}chen 2-3, D-07745 Jena, Germany
}

\date{Accepted XXX. Received YYY; in original form ZZZ}

\pubyear{2022}

\begin{document}
\label{firstpage}
\pagerange{\pageref{firstpage}--\pageref{lastpage}}
\maketitle

\begin{abstract}
When considering gaps in debris discs, a typical approach is to invoke clearing by an unseen planet within the gap, and derive the planet mass using Wisdom overlap or Hill radius arguments. However, this approach can be invalid if the disc is massive, because this clearing would also cause planet migration. This could result in a calculated planet mass that is incompatible with the inferred disc mass, because the predicted planet would in reality be too small to carve the gap without significant migration. We investigate the gap that a single embedded planet would carve in a massive debris disc. We show that a degeneracy is introduced, whereby an observed gap could be carved by \textit{two} different planets: either a  high-mass, barely-migrating planet, or a smaller planet that clears debris as it migrates. We find that, depending on disc mass, there is a minimum possible gap width that an embedded planet could carve (because smaller planets, rather than carving a smaller gap, would actually migrate through the disc and clear a wider region). We provide simple formulae for the planet-to-debris disc mass ratio at which planet migration becomes important, the gap width that an embedded planet would carve in a massive debris disc, and the interaction timescale. We also apply our results to various systems, and in particular show that the disc of ${\HD107146}$ can be reasonably well-reproduced with a migrating, embedded planet. Finally, we discuss the importance of planet-debris disc interactions as a tool for constraining debris disc masses.
\end{abstract}


\section{Introduction}

Over recent years, broad gaps have been observed in several debris discs. These systems include ${\HD107146}$ \citep{Ricci2015}, ${\HD92945}$ \citep{Marino2019}, ${\HD15115}$ \citep{MacGregor2019}, and ${\HD206893}$ \citep{Marino2020}, where gaps have been revealed at millimetre wavelengths by ALMA. Gaps are also seen in scattered light for a number of systems, including ${\HD131835}$ \citep{Feldt2017}, ${\HD120326}$ \citep{Bonnefoy2017}, and ${\HD141943}$ \citep{Boccaletti2019}. Such gaps are often ascribed to unseen planets, which can perturb and remove debris from the gap in a number of ways; commonly-considered mechanisms include a single, non-migrating planet located within the gap that scatters debris from some chaotic zone surrounding its orbit \citep{Wisdom1980, Mustill2012, Morrison2015, Nesvold2015}, or multiple planets spanning the gap clear the region of debris \citep{Faber2007, Shannon2016, Lazzoni2018}, or a precessing planet located interior to the disc carves a gap further out through secular effects \citep{Pearce2015, Yelverton2018, Sefilian2021}. Of these processes, the first two (scattering by some \textit{in situ} planet or planets) are the most intuitive, and are therefore the most commonly employed to constrain the parameters of unseen planet(s) from debris disc gaps.

However, caution is required when considering these scattering scenarios, because they do not account for the mass of the debris disc. This is only safe if the inferred planet is significantly more massive than the disc, because the dynamical effect of such a disc on the planet would be negligible. However, for smaller planets whose masses are not significantly greater than the disc, this assumption of non-evolving planets carving a gap \textit{in situ} no longer holds. This is a particular problem if the gap is narrow, because the planets predicted by the above scattering theories would have low masses. More broadly, it is an issue for any debris disc, because debris disc masses are very uncertain and estimates differ by orders of magnitude (e.g. \citealt{Krivov2021}); other than the Asteroid and Kuiper Belts (where we directly observe the large bodies comprising most of their mass), we cannot be sure that \textit{any} debris disc is significantly less massive than the planet(s) we infer from it.

In reality, any planet(s) embedded in a debris disc would migrate as they scattered particles \citep{Ida2000,Kirsh2009, Pearce2014, Pearce2015}, which would bring the planet(s) into contact with more particles to scatter, and potentially prolong migration. The result would be a wider gap than that inferred by studies considering only a massless disc. In this paper, we consider the gap carved by a single planet embedded in a debris disc, when the back-reaction of massive debris particles on the planet is included. We first consider the debris-disc mass required to drive significant planet migration (Section \ref{sec: migrationCriterion}), and provide a simple analytic argument to identify when this effect is important (Equation \ref{eq: discMassForMigration}). We then consider the gap resulting from such an interaction (Section \ref{sec: gapCarvedByMigratingPlanet}), providing simple expressions for the gap width carved when the debris mass is non-negligible (Equation \ref{eq: gapWidthApprox}), and the minimum gap width possible for a massive debris disc if the gap is carved by a single, embedded planet (Equation \ref{eq: minGapWidth}). In Section \ref{subsec: interactionTimescales} we quantify the interaction timescales (Equation \ref{eq: interactionTimescale}). We discuss our results in Section \ref{sec: discussion}, including the complicating effects of resonance sweeping, applications to known debris discs (particularly ${\HD107146}$), \textit{caveats} of our modelling, and how our results could yield insights into debris disc masses, unseen planet properties, and system evolution histories. We conclude in Section \ref{sec: conclusions}.

\section{Interaction analysis}
\label{sec: analyticPredictions}

We consider a scenario where a planet on a circular orbit is initially embedded in a `sea' of massive debris particles (the physical feasibility of this scenario is discussed in Section \ref{subsec: caveats}). This planet would scatter nearby debris, which would have two effects; a gap would open up around the planet, and the back-reaction of the disc would cause the planet to migrate. The outcome of this interaction can be found by considering energy arguments. A debris particle near the planet can be scattered either outwards or inwards, and in doing so would gain or lose energy, respectively. It would later pass close to the planet again, and get scattered again. Over time, there is a general trend for the particle to be scattered outwards, and eventually be ejected from the system. This net gain in particle energy causes the planet to lose energy, such that it becomes more tightly bound and its semimajor axis decreases. Barring a few specific cases, the typical effect of scattering in a single-planet system is therefore that the planet moves closer to the star (e.g. \citealt{Kirsh2009})\footnote{Outward migration is more easily achieved in multi-planet systems, because an outer planet can scatter material inwards where it is then ejected by an interior planet. This process is thought to have driven early outward migration of Saturn, Uranus and Neptune in the Solar System \citep{Tsiganis2005}.}.

As the planet moves inwards, it encounters and scatters more debris. This causes the planet to move in further. However, the planet would not necessarily migrate all the way to the disc inner edge; there may not be enough energy available in the disc to drive the planet all the way inwards, in which case migration would eventually stall. This is perhaps clearer if we consider an extreme case; if we dropped Jupiter into a broad disc with a very low mass, then the planet would eject material immediately around it (causing it to migrate inwards a small distance), but we would not expect such a tenuous disc to be able to significantly alter Jupiter's semimajor axis. In this case the migration of Jupiter would quickly stall, simply because the disc lacks the orbital energy required to drive Jupiter all the way to the disc inner edge. However, if we increased the disc mass then our hypothetical Jupiter would migrate inwards a bit more before stalling, and eventually, if we increased the disc mass enough, then the planet could migrate all the way to (or even beyond) the disc inner edge. We can therefore use energy arguments to calculate how far a planet embedded in a debris disc would migrate before stopping.

\begin{figure}
    \centering
    \includegraphics[width=8cm]{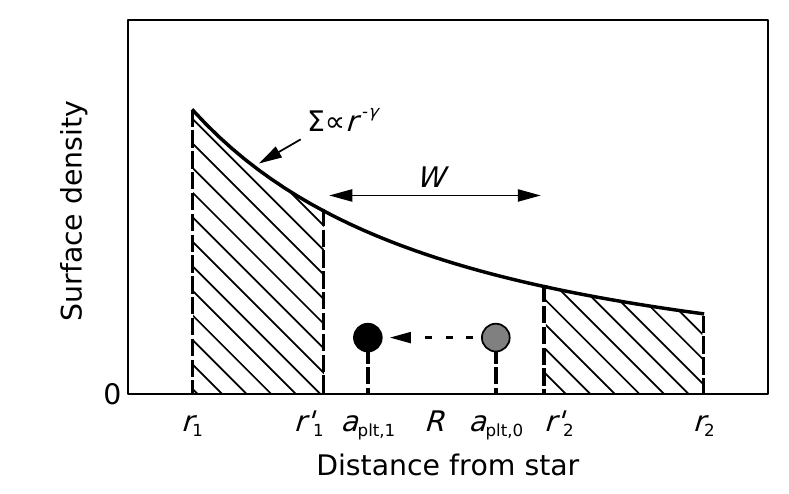}
    \caption{Schematic of the scenario we consider. A planet (grey point) at initial semimajor axis $a_{\rm plt,0}$ is embedded in a continuous debris disc, where the latter has a surface density varying with radius as ${\Sigma\propto r^{-\gamma}}$ (solid line). The disc initially spans distances $r_1$ to $r_2$ from the star. As the planet scatters material it migrates inwards, eventually coming to a stop with semimajor axis $a_{\rm plt,1}$ (black point), having ejected all debris between distances $r_1^{\prime}$ and $r_2^{\prime}$. The resulting gap has width $W$ and central radius $R$, and the hatched regions show remaining debris after the interaction has finished.}
    \label{fig: cartoon}
\end{figure}

Consider the scenario shown in Figure \ref{fig: cartoon}, where a debris disc of initial mass $M_{\rm disc}$ spans distances $r_1$ to $r_2$ from a star, and has surface density varying with distance $r$ as ${\Sigma\propto r^{-\gamma}}$ (where $\gamma$ is a constant). Embedded in this disc is a planet of mass $M_{\rm plt}$, with initial semimajor axis $a_{\rm plt,0}$. \citet{Pearce2014} showed that, if the planet \textit{just} ejects all debris between distances $r_1^\prime$ and $r_2^\prime$ from the star, then the planet will migrate inwards to a final semimajor axis $a_{\rm plt,1}$:

\begin{equation}
    \frac{a_{\rm plt,0}}{a_{\rm plt,1}} =1+\frac{M_{\rm disc}}{M_{\rm plt}}\Gamma ,
    \label{eq: migrationRatio}
\end{equation}

\noindent where

\begin{equation}
    \Gamma \equiv a_{\rm plt,0}\frac{2-\gamma}{\gamma-1}\frac{r_1^{\prime1-\gamma}-r_2^{\prime1-\gamma}}{r_2^{2-\gamma}-r_1^{2-\gamma}}
    \label{eq: gamma}
\end{equation}

\noindent if $\gamma \neq$ 1 or 2. Equations \ref{eq: migrationRatio} and \ref{eq: gamma} are valid for planets with both zero and non-zero eccentricities, but for this paper we only consider planets on circular orbits.  The equations show that the migration distance depends in some way on the disc-to-planet mass ratio, with smaller planets migrating larger distances. We will now use Equations \ref{eq: migrationRatio} and \ref{eq: gamma} to find simple expressions for whether an embedded planet would significantly migrate (Section \ref{sec: migrationCriterion}),  predict the gap width carved by a planet embedded in a massive debris disc (Section \ref{sec: gapCarvedByMigratingPlanet}), and estimate the timescales for this interaction (Section \ref{subsec: interactionTimescales}). We will also test these predictions using \mbox{$n$-body} simulations.

\subsection{Criterion for significant planet migration}
\label{sec: migrationCriterion}

We first derive a simple expression for whether a planet embedded in a massive debris disc would undergo significant migration. A planet would scatter material from a region around its orbit, and in doing so, would migrate inwards. We assume that the planet would clear some multiple $k$ Hill radii\footnote{We use Hill radii, rather than the resonance overlap criterion \citep{Wisdom1980}, to describe the chaotic region because the simpler $M{_{\rm plt}^{1/3}}$ scaling of the former is easier to work with than the $M{_{\rm plt}^{2/7}}$ scaling of the latter. It is also unclear how the \cite{Wisdom1980} criterion applies to eccentric planets whose orbits take them out of the region of overlapping resonances, whilst the Hill radius can be generalised to eccentric planet orbits \citep{Pearce2014}.} either side of its orbit; we define `significant' migration to have occurred if, in the process of clearing material, the planet migrates inwards of this region (i.e. if ${a_{\rm plt, 0} - a_{\rm plt, 1} > k R_{\rm Hill}(a_{\rm plt, 0})}$, where ${R_{\rm Hill}(a_{\rm plt}) = a_{\rm plt}[M_{\rm plt}/(3 M_*)]^{1/3}}$ is the Hill radius and $M_*$ the star mass). This criterion is justified because, if the planet scatters unstable material from a region around its initial orbit and in doing so migrates out of that region, then it will come into contact with more material, allowing it to scatter that and migrate further. We can substitute this criterion into Equations \ref{eq: migrationRatio} and \ref{eq: gamma} to derive the disc mass required for significant planet migration; assuming ${R_{\rm Hill} \ll a_{\rm plt, 0}}$, then significant migration would occur if 

\begin{equation}
\frac{M_{\rm disc}}{M_{\rm plt}} \gtrsim \frac{1}{1-k\left(\frac{M_{\rm plt}}{3M_*}\right)^{1/3}} \frac{1}{2(2-\gamma)}\frac{r_2^{2-\gamma} - r_1^{2-\gamma}}{a_{\rm plt, 0}^{2-\gamma}}.
\label{eq: discMassForMigrationAnyK}
\end{equation}

 However, the above equation depends on the arbitrary parameter $k$, the number of Hill radii cleared either side of the planet orbit. We can gain further insight by considering the infinitesimal limit where $k\rightarrow0$ (i.e. if the planet clears a very narrow region around its orbit, whether that causes the planet to migrate far enough to encounter more material, scatter that and continue migrating).  We will use \mbox{$n$-body} simulations to show that Equation \ref{eq: discMassForMigrationAnyK} with $k=0$ provides a good criterion for whether migration occurs (see below). Therefore, if a planet of mass $M_{\rm plt}$ and semimajor axis ${a_{\rm plt}}$ is embedded in a particle disc of mass ${M_{\rm disc}}$, then that planet is expected to significantly migrate inwards if

\begin{equation}
\frac{M_{\rm disc}}{M_{\rm plt}} \gtrsim \frac{1}{2(2-\gamma)}\frac{r_2^{2-\gamma} - r_1^{2-\gamma}}{a_{\rm plt}^{2-\gamma}},
\label{eq: discMassForMigration}
\end{equation}

\noindent where ${\gamma \neq 2}$ is the disc surface density index (${\Sigma \propto r^{-\gamma}}$), and $r_1$ and $r_2$ are the disc inner and outer edges respectively\footnote{Note that $M_{\rm disc}$ is the mass of the \textit{undisrupted} disc, i.e. the pre-interaction disc where the surface density index $\gamma$ is constant over the whole disc. If we instead consider the post-interaction disc, where all debris between radii $r_1^{\prime}$ and $r_2^{\prime}$ has been removed, then the mass of the gapped disc $M_{\rm disc}^\prime$ is related to the pre-interaction mass $M_{\rm disc}$ by ${M_{\rm disc}^\prime = M_{\rm disc} \,\Bigl[1-\Bigl(r_2^{\prime2-\gamma}-r_1^{\prime2-\gamma}\Bigr)/\Bigl(r_2^{2-\gamma}-r_1^{2-\gamma}\Bigr)\Bigr]}$.} (${\gamma=1.5}$ for the Minimum Mass Solar Nebula; \citealt{Weidenschilling1977, Hayashi1981}). Equation \ref{eq: discMassForMigration} does not strongly depend on $\gamma$, but if the reader requires ${\gamma=2}$, then a value close to 2 (such as 2.01) should be used instead; a similar approach can be used for ${\gamma=1}$ or 2 in Equations \ref{eq: migrationRatio} and \ref{eq: gamma}, and in any other equation containing $\gamma$. Note that Equation \ref{eq: discMassForMigration} is written in terms of total disc mass, but it is the local surface density that actually sets whether migration will occur; since the local surface density at distance $r$ is

\begin{equation}
\Sigma(r) {\rm d}r = M_{\rm disc} \frac{2-\gamma}{2\upi \left(r_2^{2-
\gamma} - r_1^{2-\gamma}\right)} r^{-\gamma} {\rm d}r,
\label{eq: surfaceDensity}
\end{equation}

\noindent we can alternatively state that significant planet migration is expected if

\begin{equation}
M_{\rm plt} \lesssim 4\upi \Sigma(a_{\rm plt}) a_{\rm plt}^2.
\label{eq: localSurfaceDensityForMigration}
\end{equation}

\noindent This is very similar to Equation 24 in \cite{Bromley2011}.

We now describe a suit of \mbox{$n$-body} simulations used to test Equation \ref{eq: discMassForMigration}, i.e. to show that Equation \ref{eq: discMassForMigrationAnyK} with $k=0$ provides a good criterion for whether significant planet migration occurs. These simulations had randomised parameters, and will also be used in later sections (we also run a suit of non-randomised simulations to model the disc of ${\HD107146}$; Section \ref{subsec: hd107146}). We simulated ${\sim100}$ different disc-planet setups with the \mbox{$n$-body} integrator {\sc rebound}, using the {\sc ias15} integrator \citep{Rein2012, ReinIAS152015}. Each simulation consisted of a central star with mass ${0.08-2\mSun}$, a planet with mass ${0.01-3000\mEarth}$ (${3\times10^{-5} - 10\mJup}$), and a debris disc with mass ${0.1-1000\mEarth}$. The disc inner and outer edges and the planet's initial semimajor axis were each drawn from the range ${0.1-1000\au}$, with the additional constraint that at least five Hill radii initially separated the planet from either disc edge. All of the above parameters were drawn for each simulation using uniform-log distributions. The initial disc surface density index was typically set to ${\gamma = 1.5}$, but was as low as -0.25 in some simulations. At least 350 massive debris particles were used to populate each disc (typically 1000), where the number of particles was chosen such that the planet would typically be at least 100 times more massive than each debris particle (in the large majority of our simulations, the planet was at least 1000 times the mass of each debris particle). For each simulation every disc particle had an initial inclination and eccentricity uniformly drawn between 0-$5^{\circ}$ and 0-0.05 respectively, and an argument of pericentre, longitude of ascending node and mean anomaly each uniformly drawn between ${0-360^{\circ}}$ (as in \citealt{Pearce2021}). Each simulation was run for at least ten diffusion timescales (or until evolution had clearly stalled), where

\begin{equation}
t_{\rm diff} \sim 0.01T_{\rm plt}\sqrt{\frac{a_{\rm plt}}{a}}\biggl(\frac{M_{\rm plt}}{M_*}\biggr)^{-2}
\label{eq:tdiff}
\end{equation}

\noindent is the diffusion timescale for a particle with semimajor axis $a$ and planet with orbital period ${T_{\rm plt}}$ \citep{Tremaine1993}; ${10 t_{\rm diff}}$ is expected to represent the scattering timescale if ${M_{\rm disc} \ll M_{\rm plt}}$ \citep{Pearce2014}. We set ${a=a_{\rm plt}}$ in the above to calculate the simulation time (in Section \ref{subsec: interactionTimescales} we consider better timescale estimates, which are also valid for ${M_{\rm disc} \gg M_{\rm plt}}$). Our simulations had a wide range of $t_{\rm diff}$ values, the majority being ${1-1000\myr}$.

\begin{figure*}
    \centering
    \includegraphics[width=17cm]{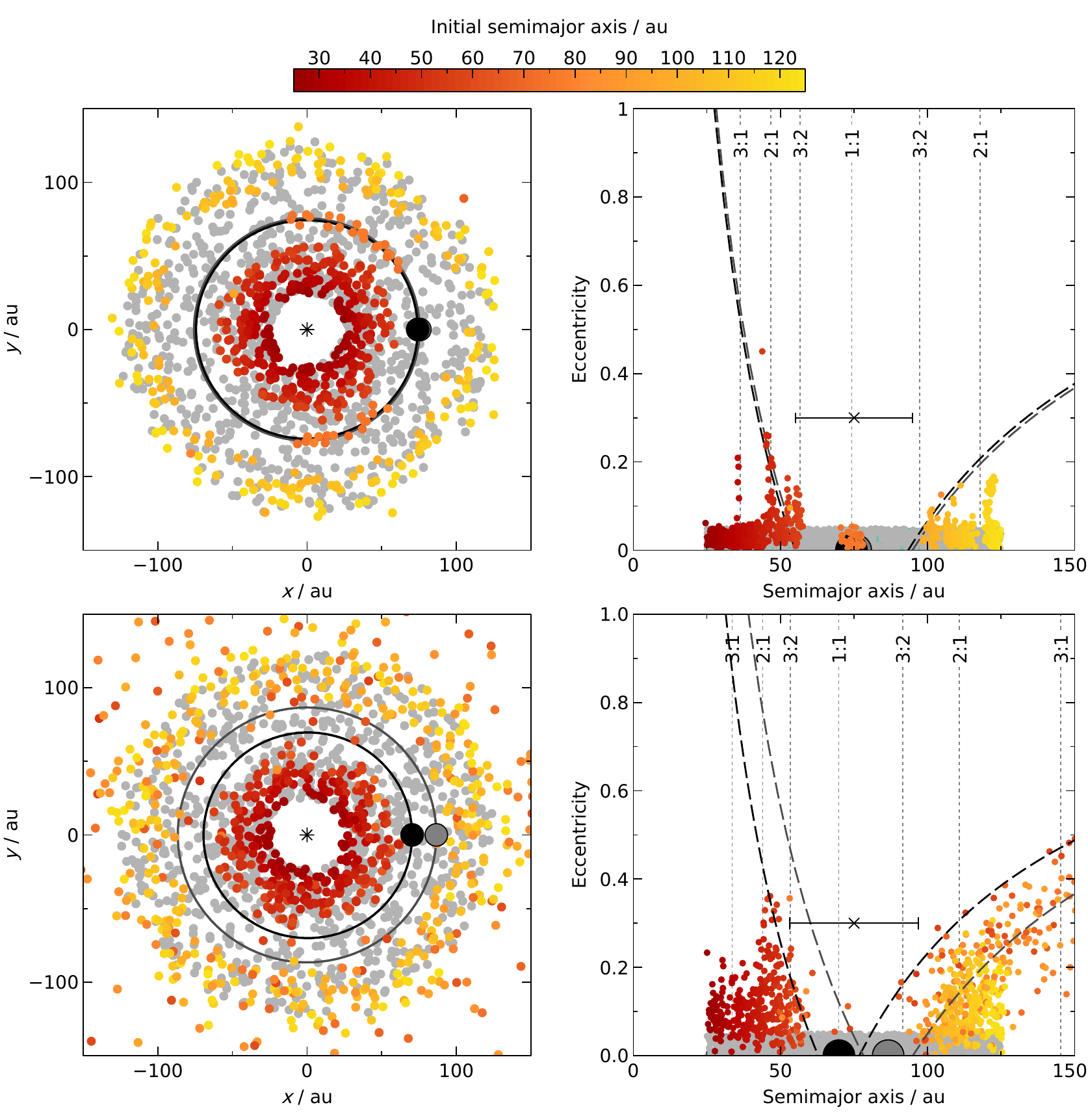}
    \caption{Similar gaps can be carved by two different planets: either a larger, barely-migrating planet (top plots), or a smaller, strongly-migrating planet (bottom plots). The figure shows two \textit{n}-body simulations with identical initial discs; the only differences between setups are the masses and initial locations of the planets, and the simulation run times. The initial discs had mass ${30\mEarth}$, span ${25-125\au}$, and surface density ${\propto r^{-1}}$, and orbit a ${1\mSun}$ star. The top and bottom plots show simulations with a $2\mJup$ and $0.1\mJup$ ($30\mEarth$) planet respectively, initially located at $76$ and $87\au$. The simulations are shown after two diffusion times have elapsed (${3.7\myr}$ for the top plots, ${1.6\gyr}$ for the bottom plots). Left plots: orbital positions of all particles after two diffusion times (coloured), and their initial positions (grey). The star is marked by the asterisk. The planet positions and orbits are marked by the large circles and lines for the initial (grey) and final (black) times, respectively. Debris particles are coloured by their initial semimajor axes. Right plots: eccentricities and semimajor axes of the planets and particles at the initial and final times. The dashed lines show particle orbits coming within ${3r_{\rm Hill}}$ of the planets' orbit at the initial (grey) and final (black) times. The crosses and bars show the predicted gap widths calculated with Equation \ref{eq: gapWidthApprox}, where we solved for $R$ numerically in Equations \ref{eq: migrationRatio} and \ref{eq: gamma} beforehand. The planets' migration over time is shown in Figure \ref{fig: planetMigration}. Note that, despite the final disc morphologies being similar in the two simulations, the debris-excitation levels and the relative scattered-disc populations differ considerably between the high- and low-mass planet scenarios (discussed in Section \ref{subsec: insightDiscussion}).
}
    \label{fig: simulationSetup}
\end{figure*}

\begin{figure}
    \centering
    \includegraphics[width=8cm]{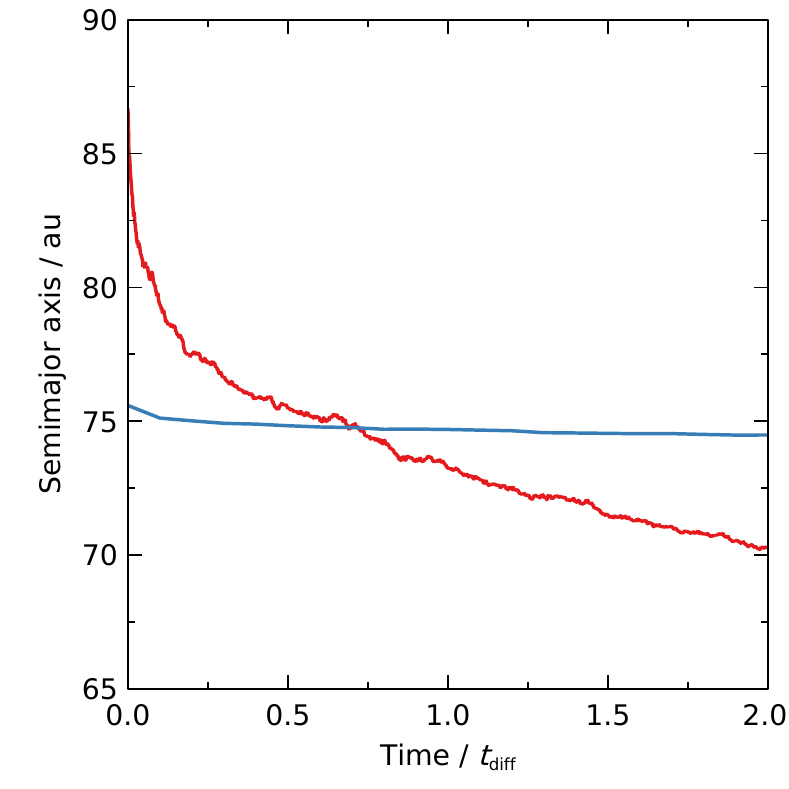}
    \caption{Semimajor axis evolution of the planets from the simulations in Figure \ref{fig: simulationSetup}, where time is expressed in diffusion times (Equation \ref{eq:tdiff}). The lines show the migration of a $2\mJup$ (blue) and $0.1\mJup$ ($30\mEarth$, red) planet, embedded in a $30\mEarth$ debris disc. Only the first two diffusion times are shown here (${3.7\myr}$ for the blue line, and ${1.6\gyr}$ for the red line), during which the majority of the planet evolution occurs.}
    \label{fig: planetMigration}
\end{figure}

Two example simulations are shown in Figures \ref{fig: simulationSetup} and \ref{fig: planetMigration}. These simulations show two cases: one with significant planet migration, and one where the planet's semimajor axis remains roughly unchanged. Gaps are carved in the debris disc in each case, and we will discuss these specific example simulations in more detail throughout the paper. In all our simulations the planet either barely migrates at all, or it migrates inwards. In all cases the planet eccentricity remains low throughout the interaction, the disc stays roughly axisymmetric, and the planet migration is smooth\footnote{In a sea of infinite particles, a scattering planet is expected to smoothly migrate whilst its eccentricity is continually damped. Since we see this in our simulations, this implies that the masses of our individual disc particles are sufficiently low to avoid stochastic effects.}.

    
\begin{figure}
    \centering
    \includegraphics[width=8cm]{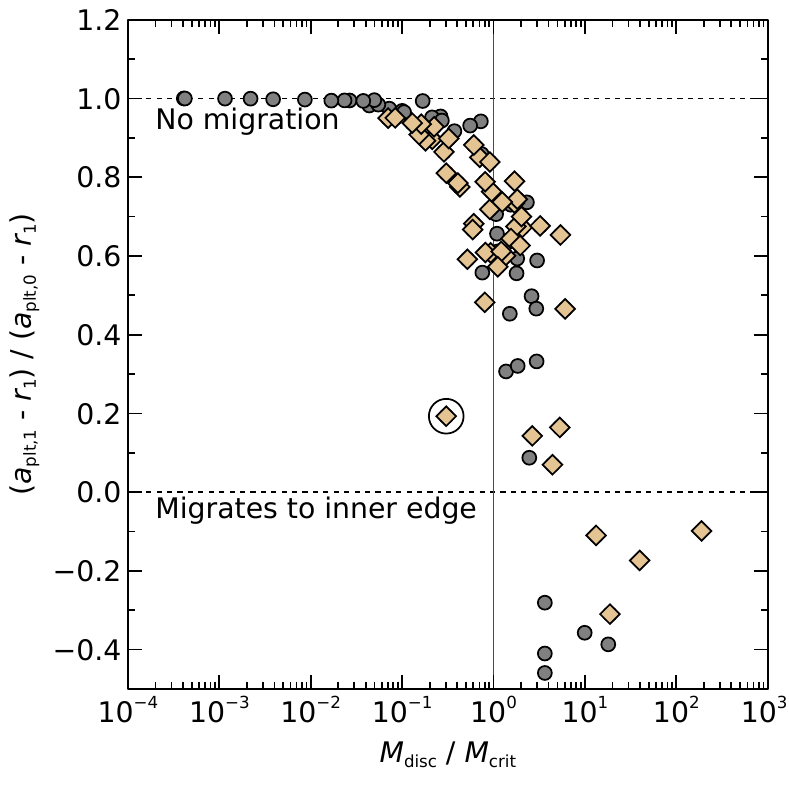}
    \caption{Comparing our analytic prediction for whether significant planet migration occurs to the simulation results.  Horizontal axis: initial debris-disc masses, in terms of the critical disc mass $M_{\rm crit}$ predicted to cause significant planet migration (Equation \ref{eq: discMassForMigration}). Simulations with ${M_{\rm disc} / M_{\rm crit} \gtrsim 1}$ are expected to result in significant planet migration. Vertical axis: a coefficient showing how far the planet actually migrated in each simulation, where a value of 1 means the planet did not migrate at all, and a value of 0 means the planet migrated to the disc inner edge (note that it is possible for planets to migrate \textit{beyond} the disc inner edge). Grey circles represent simulations with randomised initial parameters (Section \ref{sec: migrationCriterion}), and beige diamonds show targeted simulations aimed at reproducing the ${\HD107146}$ disc (Section \ref{subsec: hd107146}). The plot shows that Equation \ref{eq: discMassForMigration} provides a good prediction for whether a planet embedded in a debris disc is expected to migrate. The exception (circled) is a simulation where significant resonant interactions occurred, as discussed in Section \ref{subsec: resonances}.}
    \label{fig: MigrationPlot}
\end{figure}

 We now verify our prediction for whether significant planet migration occurs (Equation \ref{eq: discMassForMigration}). Figure \ref{fig: MigrationPlot} shows a measure of the planet-migration distances from our simulations, plotted against the disc masses. The migration coefficient (vertical axis) is defined such that planets which do not migrate have value 1, whilst planets that migrate all the way in to the disc's initial inner edge have value 0 (note it is also possible for planets to migrate \textit{beyond} the initial inner edge; in these cases the migration coefficient is negative). The disc masses (horizontal axis) are each plotted in terms of the `critical' disc mass $M_{\rm crit}$, which is our predicted disc mass for significant migration calculated for each simulation using Equation \ref{eq: discMassForMigration}. We see that Equation \ref{eq: discMassForMigration} provides a good estimate for whether significant migration occurs; in general, little planet migration occurs in simulations where the disc mass is lower than our predicted critical value (i.e. left of unity on the horizontal axis), whilst significant migration happens if the disc mass is higher than our predicted value (right of unity). If instead we had used non-zero values of $k$ in Equation \ref{eq: discMassForMigrationAnyK}, then the points would all shift to the left and our prediction would provide a worse fit to the simulations.

\subsection{Gap carved by a migrating planet}
\label{sec: gapCarvedByMigratingPlanet}

 We now consider the gap that an embedded planet would carve in a massive debris disc. We will use Equations \ref{eq: migrationRatio} and \ref{eq: gamma} to predict the gap width (Equation \ref{eq: gapWidthApprox}), and in particular show that smaller, migrating planets can carve similar gaps to larger, non-migrating ones. We will also show that there is a \textit{minimum} gap width that could be carved by a planet embedded in a massive debris disc; this can also be used to infer disc masses from gap observations.

A planet is expected to eject non-resonant particles within multiple ($k$) Hill radii either side of its orbit. For a non-migrating planet, the resulting gap would have a central radius ${R = a_{\rm plt}}$ and width ${W = 2 k h a_{\rm plt}}$, where $h$ is the dimensionless Hill radius (${h \equiv R_{\rm Hill}/a_{\rm plt}}$). Conversely, if the planet migrates inwards from semimajor axis ${a_{\rm plt,0}}$ to ${a_{\rm plt,1}}$, then the inner edge of the resulting gap would be at ${r_1' = a_{\rm plt,1} - k h a_{\rm plt, 1}}$, and the outer edge at ${r_2' = a_{\rm plt,0} + k h a_{\rm plt, 0}}$. However, the migrating planet scenario is complicated, because the planet's migration distance is set by how much material it scatters, which depends on how far it migrates. We can solve this problem by setting ${r_1'}$ and ${r_2'}$ to the values above, and substituting them into Equations \ref{eq: migrationRatio} and \ref{eq: gamma}. Defining the gap width and central radius as ${W \equiv r_2' - r_1'}$ and ${R \equiv (r_1'+r_2')/2}$ respectively (see Figure \ref{fig: cartoon}), the result is that the gap satisfies

\begin{multline}
W - 2 R k h(M_{\rm plt})  - \frac{M_{\rm disc}}{M_{\rm plt}}\frac{2-\gamma}{\gamma-1}\frac{r_2'r_1'^{2-\gamma} - r_1'r_2'^{2-\gamma}}{r_2^{2-\gamma} - r_1^{2-\gamma}} = 0,
\label{eq: gapWidthBothNotation}
\end{multline}

\noindent where the first two terms are the simple case where the planet does not migrate, and the third term is an additional planet-migration term.

The migration term in Equation \ref{eq: gapWidthBothNotation} is unwieldy, but it can be significantly simplified. Taylor-expanding Equation \ref{eq: gapWidthBothNotation} with respect to ${W / (2R)}$ (where this term is always less than unity, and often much less) yields a simpler equation for the gap width carved by a migrating planet:

\begin{equation}
W \approx \frac{2 R kh(M_{\rm plt})}{1-\frac{M_{\rm disc}}{M_{\rm plt}}(2-\gamma)\frac{R^{2-\gamma}}{r_2^{2-\gamma} - r_1^{2-\gamma}}}.
\label{eq: gapWidthApprox}
\end{equation}

\noindent This equation is valid provided ${[W / (2R)]^3 \ll 1}$, which is the case for almost all systems of interest; it will only diverge for strongly-migrating planets carving very wide gaps.  The only outstanding parameter is $k$, the number of Hill radii cleared either side of the planet's orbit; our simulations suggest that ${k\approx3}$ (see Figures \ref{fig: simulationSetup}, \ref{fig: resonanceSweeping} and \ref{fig: hd107146}), which compares well with literature definitions of encounter zones (where $k\sim3.1-3.5$; \citealt{Gladman1993, Ida2000, Kirsh2009}). Equation \ref{eq: gapWidthApprox} with ${k\approx3}$ can therefore be used to predict the width of a gap carved by a planet embedded in a massive debris disc\footnote{Note the similarity between the denominator of Equation \ref{eq: gapWidthApprox}, which quantifies when the gap width significantly deviates from the zero-mass-disc case, and Equation \ref{eq: discMassForMigration}, which shows when planet migration becomes significant.}. The equation can also be inverted and solved numerically to find the planet mass required to carve an observed gap; a {\sc python} program to do this is publicly available for download\footnote{\label{footnote: publicCode}tdpearce.uk/public-code}.

\begin{figure}
    \centering
    \includegraphics[width=8cm]{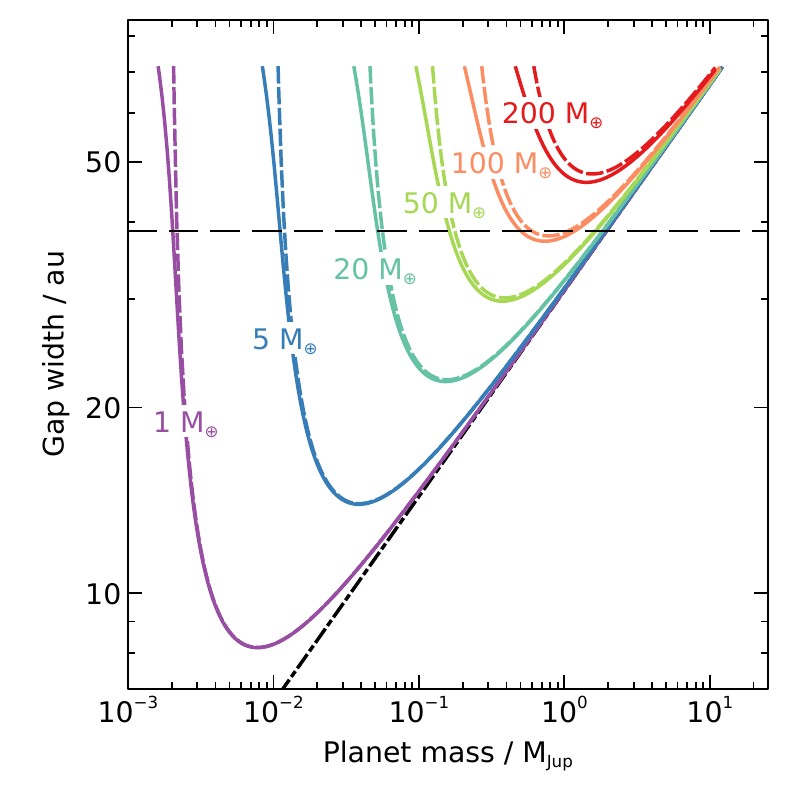}
    \caption{Predicted widths of the gaps that an embedded planet would carve in a debris disc, as functions of the planet mass and initial disc mass. For these predictions the star has mass $1\mSun$, and the disc has initial span ${40-146\au}$ and surface density $r^{-1/4}$ (which we show can replicate the morphology of $\HD107146$ reasonably well: Section 3.2.1). The gap is centred on $R=76\au$. The solid lines show the full Equation \ref{eq: gapWidthBothNotation} (evaluated numerically) for $k=3$ and ${\gamma=0.25}$, and the dashed lines the analytic approximation from Equation \ref{eq: gapWidthApprox}. Different colours show different initial disc masses, with the masses labelled in the figure. The horizontal dashed line shows ${39\au}$, the width of the gap observed in the ${\HD107146}$ disc \citep{Marino2018}. The plot shows that a given gap width can generally be carved by either a low-mass, strongly-migrating planet, or a higher mass, barely-migrating planet. For high planet masses, planet migration becomes negligible and the gap width tends to that for a zero-mass disc (${6r_{\rm Hill}}$: black dot-dash line). The {\sc python} program used to produce this plot is publicly available for download\textsuperscript{\ref{footnote: publicCode}}.}
    \label{fig: gapWidths}
\end{figure}

Figure \ref{fig: gapWidths} shows the gap width predicted from Equations \ref{eq: gapWidthBothNotation} and \ref{eq: gapWidthApprox} for an example disc, as functions of planet and disc masses. The plot shows that the same gap can be carved by \textit{two} different planets: either a high-mass, barely-migrating planet, or a low-mass, strongly-migrating planet. The inclusion of disc mass means that the gap width does not monotonically decrease with decreasing planet mass; if the planet is too small, then instead of carving a very narrow gap, it would instead strongly migrate and carve a wider gap. This leads to the counterintuitive result that a low-mass planet may carve a wider gap than a high-mass planet could\footnote{A similar effect also occurs in secular interactions; non-zero disc mass can cause low-mass planets to precess, which results in a broader unstable region \citep{Pearce2015}.}.

\begin{figure}
    \centering
    \includegraphics[width=8cm]{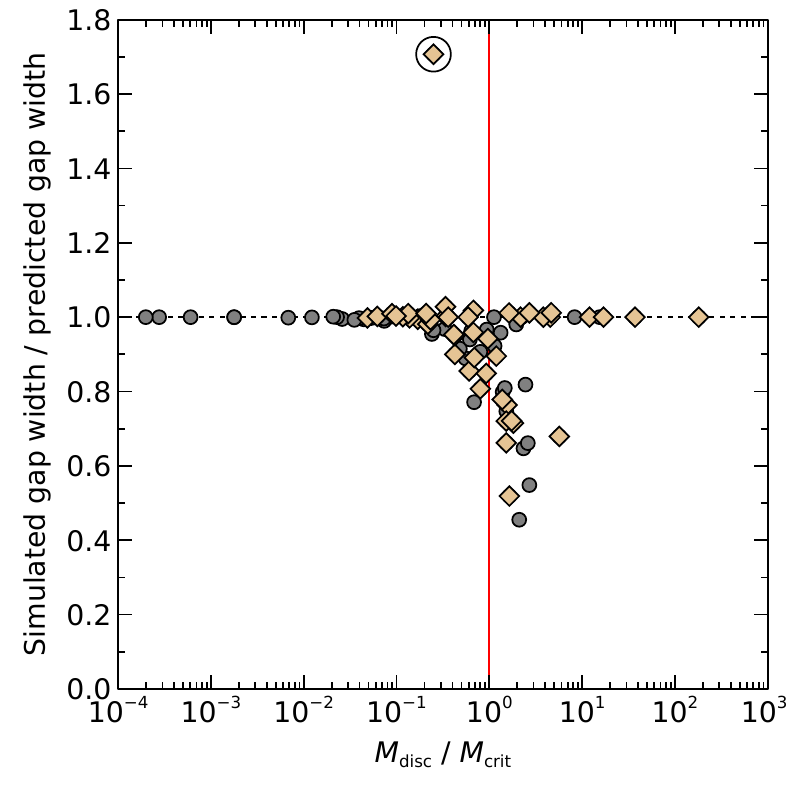}
    \caption{Comparison of the gap widths measured from our $n$-body simulations to those predicted by Equation \ref{eq: gapWidthBothNotation}. The horizontal axis shows initial disc mass in terms of $M_{\rm crit}$, the minimal disc mass at which planet migration is expected (Equation \ref{eq: discMassForMigration}). Symbols were defined in Figure \ref{fig: MigrationPlot}. The plot shows that Equation \ref{eq: gapWidthBothNotation} works reasonably well; differences between simulated and predicted gap widths at ${M_{\rm disc} / M_{\rm crit} \sim 1}$ occur because the moderate planet migration is complicated by additional effects such as resonance sweeping and incomplete debris clearing (Sections \ref{subsec: resonances} and \ref{subsec: caveats}). The plot is similar if the analytic prediction for gap width (Equation \ref{eq: gapWidthApprox}) is used instead. The circled simulation is marked on other plots, and shows a simulation where significant resonant interactions occurred (Section \ref{subsec: resonances}).}
    \label{fig: widthsVsPrediction}
\end{figure}

Figure \ref{fig: widthsVsPrediction} compares the gap widths from our $n$-body simulations to those predicted by numerically solving Equation \ref{eq: gapWidthBothNotation}. For this plot we defined the $n$-body gap widths as being ${3 r_{\rm Hill}}$ either side of the planet's initial and final semimajor axes, i.e. ${W_{\rm sim} \equiv a_{\rm plt,0}(1+3h) - a_{\rm plt,1}(1-3h)}$, and we predicted $R$ for Equation \ref{eq: gapWidthBothNotation} by solving Equations \ref{eq: migrationRatio} and \ref{eq: gamma} numerically. For both predicted and simulated gaps, if the inner edge of the gap would be interior to the initial inner edge of the disc then we truncated the gap at this radius, i.e. ${W_{\rm max} \equiv a_{\rm plt,0}(1+3h) - r_1}$. The plot shows that Equation \ref{eq: gapWidthBothNotation} works reasonably well; for very strong or very weak planet migration the agreement is \mbox{$\sim$1:1}. There is some spread for intermediate cases (${M_{\rm disc} \sim M_{\rm crit}}$, where $M_{\rm crit}$ is the minimal disc mass at which migration is expected from Equation \ref{eq: discMassForMigration}), because these interactions are complicated by additional effects such as resonance sweeping and incomplete debris clearing (Sections \ref{subsec: resonances} and \ref{subsec: caveats}). Nonetheless, almost all of the simulated gap widths agree with our predictions to within ${50\percent}$. The plot is also similar if, rather than predicting gap widths by numerically solving Equation \ref{eq: gapWidthBothNotation}, we instead used the analytic approximation from Equation \ref{eq: gapWidthApprox}.

Another prediction from Equations \ref{eq: gapWidthBothNotation} and \ref{eq: gapWidthApprox} is that a non-zero disc mass results in a \textit{minimum} gap width that can be carved by an embedded planet. This is visible as the turning points in \mbox{Figure \ref{fig: gapWidths}}. To determine this minimum width, we differentiate Equation \ref{eq: gapWidthApprox} with respect to $M_{\rm plt}$ and set the result equal to zero. This yields the minimum possible gap width that \textit{any} embedded planet could carve as

\begin{multline}
W_{\rm min} \approx 2.94 k \left[(2-\gamma) \frac{M_{\rm disc}}{M_*} \frac{R^{5-\gamma}}{r_2^{2-\gamma} - r_1^{2-\gamma}}\right]^{1/3},
\label{eq: minGapWidth}
\end{multline}


\noindent where again $k\approx3$ (note, however, that resonant sweeping can also be important here, as discussed in Section \ref{subsec: resonances}; it is possible for a narrower gap to be present, if material at the inner edge is in resonance). The planet responsible for carving this minimum-width gap would have mass

\begin{equation}
M_{\rm plt} = M_{\rm disc} \frac{4 (2-\gamma) R^{2-\gamma}}{r_2^{2-\gamma} - r_1^{2-\gamma}}.
\label{eq: pltMassForMinGapWidth}
\end{equation}

We can also invert Equation \ref{eq: minGapWidth} to gain information about the disc mass. Since ${W\geq W_{\rm min}}$, rearranging Equation \ref{eq: minGapWidth} shows that, if a given gap has been carved by an embedded planet, then there is a maximum possible mass that the initial disc could have had. This maximum mass is

\begin{equation}
M_{\rm disc} \lesssim \frac{M_* W^3}{25.4 k^3 (2-\gamma)} \frac{r_2^{2-\gamma} - r_1^{2-\gamma}}{R^{5-\gamma}}.
\label{eq: maxDiscMassForGap}
\end{equation}

\noindent So if Equation \ref{eq: maxDiscMassForGap} is not satisfied, then an observed gap \textit{cannot} have been carved by a single planet embedded in an initially continuous debris disc, unless the interaction is still ongoing (Section \ref{subsec: otherSystems}) or the disc edges are dominated by resonances (Section \ref{subsec: resonances}). Equation \ref{eq: maxDiscMassForGap} could therefore be used to gain insight into debris disc masses (Sections \ref{subsec: otherSystems} and \ref{subsec: insightDiscussion}).

Finally, in some cases it is possible for the planet to migrate all the way to the disc inner edge. We can find the regime where this occurs by solving Equations \ref{eq: migrationRatio} and \ref{eq: gamma} for ${a_{\rm plt,1} \lesssim r_1}$; making the approximation ${a_{\rm plt, 0} \approx r_{2}^\prime}$, we find that the planet can migrate all the way to the disc inner edge provided

\begin{equation}
\frac{M_{\rm disc}}{M_{\rm plt}} \gtrsim \frac{\gamma-1}{2-\gamma} \left(1-\frac{r_1}{a_{\rm plt,0}}\right) \frac{\left(r_2 / r_1 \right)^{2-\gamma}-1}{1-\left(r_1 / a_{\rm plt,0}\right)^{\gamma-1}}.
\label{eq: migrationToInnerEdge}
\end{equation}

\noindent We discuss this scenario further in Section \ref{subsec: insightDiscussion}.

\subsection{Interaction timescales}
\label{subsec: interactionTimescales}

We now consider the interaction timescale. \cite{Pearce2014} argued that, if the planet is much more massive than the disc, then the debris-scattering timescale is proportional to the diffusion time ${t_{\rm diff}}$ (Equation \ref{eq:tdiff}). However, that equation does not depend on disc mass, and since disc mass affects the planet migration rate, Equation \ref{eq:tdiff} might not hold if the planet is less massive than the disc. We will show that a simple modification of Equation \ref{eq:tdiff} to account for disc mass can well-characterise the interaction timescale.

The diffusion time  ${t_{\rm diff}}$ (Equation \ref{eq:tdiff}) depends strongly on planet mass, suggesting that planet mass drives the interaction if ${M_{\rm plt} \gg M_{\rm disc}}$. Conversely, if ${M_{\rm plt} \ll M_{\rm disc}}$, then a reasonable suggestion is that the disc mass drives the interaction instead. In the latter regime, a simple approximation would be to replace the planet mass in Equation \ref{eq:tdiff} with the disc mass; i.e. if ${M_{\rm plt} \gg M_{\rm disc}}$ then the interaction timescale depends on ${t_{\rm diff}}$ from Equation \ref{eq:tdiff}, but if ${M_{\rm plt} \ll M_{\rm disc}}$ then the timescale depends on ${t_{\rm diff} \times (M_{\rm disc}/M_{\rm plt})^{-2}}$. These regimes can be combined, such that the general interaction timescale  $t_{\rm int}$ is proportional to 

\begin{equation}
t_{\rm int} \propto 0.01 T_{\rm plt} \left( \frac{M_*}{M_{\rm plt} + M_{\rm disc}} \right)^2
\label{eq: interactionTimescale}
\end{equation}

\noindent (where we have set ${a_{\rm plt} \approx a}$ in Equation \ref{eq:tdiff}).

\begin{figure}
    \centering
    \includegraphics[width=8cm]{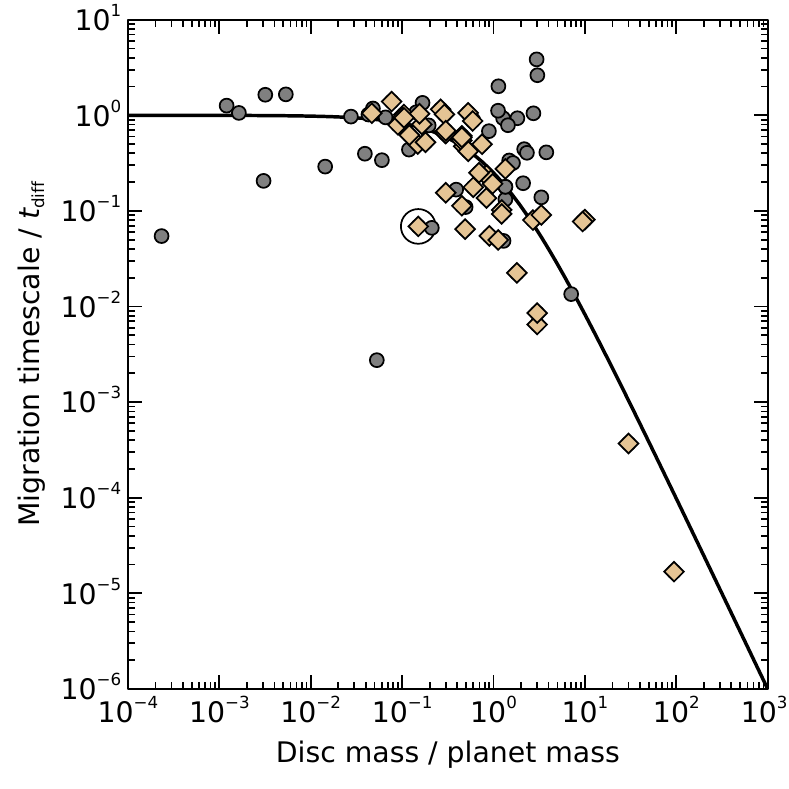}
    \caption{Planet-migration timescales from our simulations, compared to our theoretical prediction (Equation \ref{eq: interactionTimescale}). Symbols were defined in Figure \ref{fig: MigrationPlot}. The horizontal axis shows the ratio of the disc and planet masses, and the vertical axis shows the ratio of the planet-migration timescale inferred from simulations to the diffusion timescale ${t_{\rm diff}}$ (Equation \ref{eq:tdiff}). Equation \ref{eq: interactionTimescale} is shown by the solid line. The interaction is driven by the larger of the disc and planet masses.}
    \label{fig: migrationTimescale}
\end{figure}

Figure \ref{fig: migrationTimescale} shows that Equation \ref{eq: interactionTimescale} provides a good estimate for the planet-migration timescale. The figure shows the planet-migration timescales from our simulations (defined as the time it takes the planet to migrate ${70\percent}$ of its total migration distance, where we only include simulations that have run for at least ${t_{\rm diff}}$ whose planet migration has clearly stalled). We show the ratio of this migration timescale to ${t_{\rm diff}}$ (Equation \ref{eq:tdiff}), versus the ratio of the disc and planet masses. For planets more massive than the disc, the migration timescale is close to ${t_{\rm diff}}$; conversely, for planets less massive than the disc, the migration timescale has an additional ${(M_{\rm disc}/M_{\rm plt})^{-2}}$ dependence. The line shows our prediction from Equation \ref{eq: interactionTimescale}, which provides a good fit. The general planet-migration timescale can therefore be approximated by Equation \ref{eq: interactionTimescale}. A second timescale, that for most unstable debris to be ejected, will be longer; \cite{Pearce2014} found this took around ten diffusion times if ${M_{\rm plt} \gg M_{\rm disc}}$, and our simulations suggest a similar scaling with our Equation \ref{eq: interactionTimescale} if ${M_{\rm plt} \ll M_{\rm disc}}$. In summary, for the scenario considered here, the planet-migration timescale is roughly Equation \ref{eq: interactionTimescale}, whilst the timescale for debris clearing (and thus deep gap creation) is about ten times this.

\section{Discussion}
\label{sec: discussion}

Having characterised the interaction and made analytical predictions for its outcome, we now discuss several further considerations and potential applications of our work. In Section \ref{subsec: resonances} we discuss an additional effect seen in our simulations: debris being swept up into mean-motion resonances. In Section \ref{subsec: applicationToObservedDebrisDiscs} we apply our results to several observed debris discs, including the ${\HD107146}$ and Solar-System discs. Section \ref{subsec: caveats} describes several \textit{caveats} that must be considered when applying our results, and Section \ref{subsec: insightDiscussion} discusses potential wider insights into disc and system properties that could be found from our work.

\subsection{Resonance sweeping affects the gap inner edge}
\label{subsec: resonances}

Our analysis assumed scattering to be the dominant dynamical interaction in this scenario. However, another effect manifests itself in many of our simulations: resonance sweeping. This occurs when a planet migrates, causing the nominal locations of its associated mean-motion resonances to migrate as well. If these resonance locations move through debris, then particles can be `swept up' and trapped, resulting in significant populations of resonant bodies \citep{Wyatt2003}.

\begin{figure}
    \centering
    \includegraphics[width = 8cm]{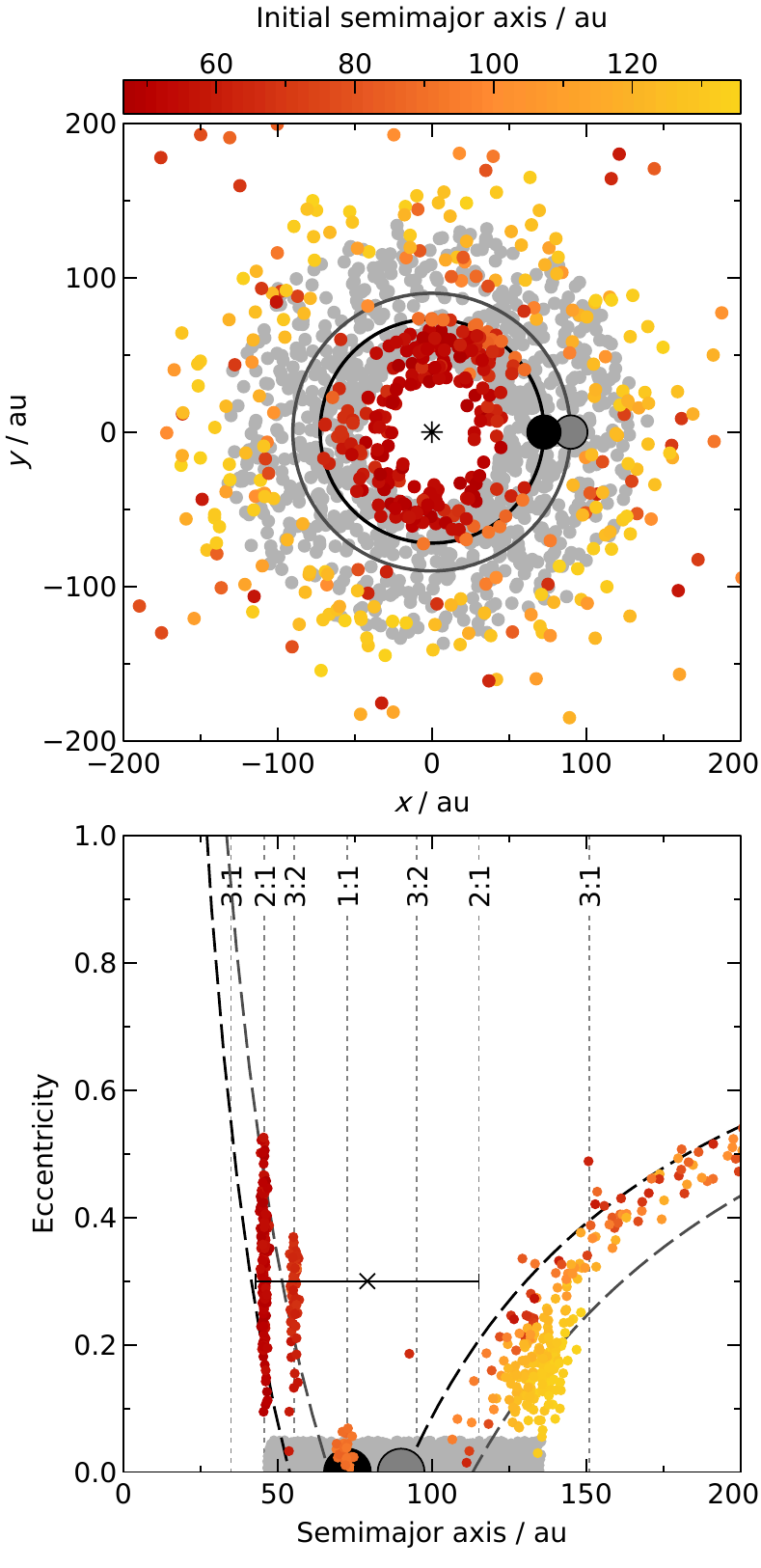}
     \caption{A simulation showing significant resonance sweeping by a migrating planet. Symbols were defined in Figure \ref{fig: simulationSetup}. The simulation setup had a $2\mJup$ planet initially located at $90\au$, embedded in a disc with initial mass $300\mEarth$, span $47-136\au$, and surface density $\propto r^{-1.5}$. The Figure shows the simulation after $6.8\myr$, over which time the planet migrated inwards by $17\au$ and swept significant quantities of debris into the 2:1 and 3:2 resonances. Top: orbital position of all particles. Bottom: eccentricity and semimajor axis of the planet and particles. From our analytic prediction for gap width (Equation \ref{eq: gapWidthApprox}, horizontal bar on lower plot), it was expected that all debris initially interior to the planet would be ejected, and hence the migrating planet would truncate the disc (rather than form a gap). However, in this case material was swept into resonances and protected from ejection (Section \ref{subsec: resonances}); this resulted in a gapped-disc morphology, where the inner edge is entirely resonant.}
    \label{fig: resonanceSweeping}
\end{figure}

Resonance sweeping has three primary effects in our scenario. Firstly, if material initially interior to the planet is swept into resonance as the planet migrates inwards, then that material can be protected from later scattering by the planet. This is because the resonances migrate inwards with the planet; material trapped in resonance therefore also moves inwards, and many of these particles maintain orbital configurations that never come close enough to the planet to be scattered. An example of this is shown in Figure \ref{fig: resonanceSweeping}. Here, some of the material initially interior to the planet has been swept up into the 2:1 and 3:2 resonances (bottom plot) and pushed inwards, and is protected from scattering. All non-resonant material initially interior to the planet has been ejected, so in this case a gap forms where the inner edge is entirely resonant. This means that, even if our analysis suggests that a planet would migrate in and scatter all material initially interior to it, it is still possible for a gap to be formed where the inner edge of the disc is resonant. Resonant material at the gap inner edge may have a pronounced non-axisymmetric structure (top plot); however, this particular simulation is an extreme case, and often the particle eccentricities are lower and any resonant structure is much less pronounced (e.g. Figure \ref{fig: hd107146}).

Secondly, resonance sweeping can affect how far the planet migrates, and this migration distance (and the resulting gap width) can be either larger or smaller than it would be under scattering alone. In most of our simulations, resonances caused planets to migrate less far than we predicted. This occurred for two reasons: first, because some of the material that the planet would otherwise scatter is instead swept up into resonance and `protected', so the planet scatters less material and its migration stalls prematurely, and second, because the migrating planet pushes resonant debris inwards too, so some of the energy transferred in the scattering process is delegated to resonant planetesimals instead of the planet. However, some of our simulations showed the opposite effect: the planet migrated further than it would through scattering alone. This occurred because material in resonance librates in eccentricity, and the maximum eccentricity a resonant particle reaches grows as the planet migrates inwards. An example of such eccentricity libration can be seen in Figure \ref{fig: resonanceSweeping}, where resonant material interior to the planet now periodically cycles up to high eccentricities (0.4-0.6 in this case). If the maximum eccentricities increase high enough, then particle orbits can evolve such that resonant particles may pass close to the planet and be scattered, despite the planet and particle semimajor axes differing significantly. This process allows the planet to come into contact with material well-interior to its orbit, which it would otherwise be unable to scatter (e.g. \citealt{Reche2008}). By ejecting resonant material with significantly lower semimajor axes than it, the planet can loose considerable energy, and migrate further inwards than it otherwise would. This process occurred in the circled simulation in Figures \ref{fig: MigrationPlot}, \ref{fig: widthsVsPrediction} and \ref{fig: migrationTimescale}, allowing that planet to migrate significantly further inwards than our scattering model implied.

Thirdly, resonance sweeping excites debris, and increases the dynamical temperature of the disc. Figure \ref{fig: resonanceSweeping} shows this; yellow particles with semimajor axes of ${125-135\au}$ and eccentricities ${\approx 0.1}$ could not have reached these orbits through scattering alone, since they lie below the dashed lines denoting orbits that could come within 3 Hill radii of the planet. Instead, these particles were excited as resonances (particularly the 2:1 resonance) swept through the disc as the planet migrated inwards. This effect could potentially stir an initially cold debris disc, igniting a collisional cascade that produces observable dust. This excitation is also apparent in Figure \ref{fig: simulationSetup}; the disc exterior to the barely-migrating planet (top plots) has a low excitation level, whilst in the case where the planet strongly migrates (bottom plots), the equivalent debris has considerably higher eccentricities.

\subsection{Application to observed debris discs}
\label{subsec: applicationToObservedDebrisDiscs}

We now apply our analyses to several debris disc systems as examples. We start with a detailed analysis of ${\HD107146}$, before considering ${\HD15115}$, ${\HD92945}$, ${\HD206893}$ and the Solar System.

\subsubsection{${\HD107146}$}
\label{subsec: hd107146}

${\HD107146}$ is an approximately ${150\myr}$ old G2V star, hosting an ALMA-resolved debris disc spanning ${47-136\au}$ with a \mbox{$39\au$-wide} `gap' centred at ${76\au}$ \citep{Ricci2015, Marino2018}. The surface brightness of this gap is lower than the surrounding disc, but still non-zero; material clearly exists in the gap itself. Various theories have been proposed to explain this feature, including secular interactions with a planet interior to the disc inner edge \citep{Pearce2015, Yelverton2018, Sefilian2021}, or debris scattering by one or more non-migrating planets located in the gap \citep{Ricci2015, Marino2018}. Models of a single, non-migrating planet carving the gap \textit{in situ} struggle to explain observations, because if such a planet were massive enough to carve the wide gap, it would deplete too much debris to be compatible with observations. However, such models neglect the disc mass; since this could be ${\gtrsim 100\mEarth}$ \citep{Ricci2015, Marino2018}, any embedded planets up to Jovian masses would undergo significant migration. In this section we show that the ${\HD107146}$ disc can be reproduced reasonably well if a single planet has migrated through the massive disc.

Equation \ref{eq: maxDiscMassForGap} shows that, for the gap to be carved by a single embedded planet, the initial disc mass in non-planetary material would have to have been ${\lesssim 100\mEarth}$ (this holds for a wide range of initial surface density indices $\gamma$). This can also be seen on \mbox{Figure \ref{fig: gapWidths}}; the plot shows the predicted gap width as a function of planet and disc masses for a gap centred on ${76\au}$ in the ${\HD107146}$ disc, and a gap of width ${39\au}$ could not have been carved through scattering by a single embedded planet if the initial disc mass were greater than ${100\mEarth}$ (note that the plot shows a slightly wider disc initially spanning ${40-146\au}$ with surface density ${\propto r^{-1/4}}$, for consistency with the analysis below). This disc mass is smaller than other estimates, but recall that estimating debris disc masses is very uncertain and prone to large uncertainties (e.g. \citealt{Krivov2021}). If the initial disc mass were ${50\mEarth}$, then solving Equation \ref{eq: gapWidthApprox} with ${\gamma=1/4}$ (assuming the planet clears ${k=3}$ Hill radii either side of it) suggests that the ${39\au}$ gap could be carved either by a planet with mass ${M_{\rm plt} = 0.16\mJup}$ migrating inwards from 85 to ${63\au}$, or a ${1.6 \mJup}$  planet migrating inwards from 76 to ${74\au}$ (as shown in Figure \ref{fig: gapWidths}). Again, these masses do not depend strongly on $\gamma$; if instead ${\gamma=1.5}$, then Equation \ref{eq: gapWidthApprox} predicts ${M_{\rm plt} = 0.23}$ or ${1.6 \mJup}$. However, these predictions assume that the planet ejects all material it encounters; since the gap of ${\HD107146}$ is not fully depleted of material, we will find that this particular system is better reproduced if the planet is smaller than these values.

\begin{figure*}
    \centering
    \includegraphics[width=17cm]{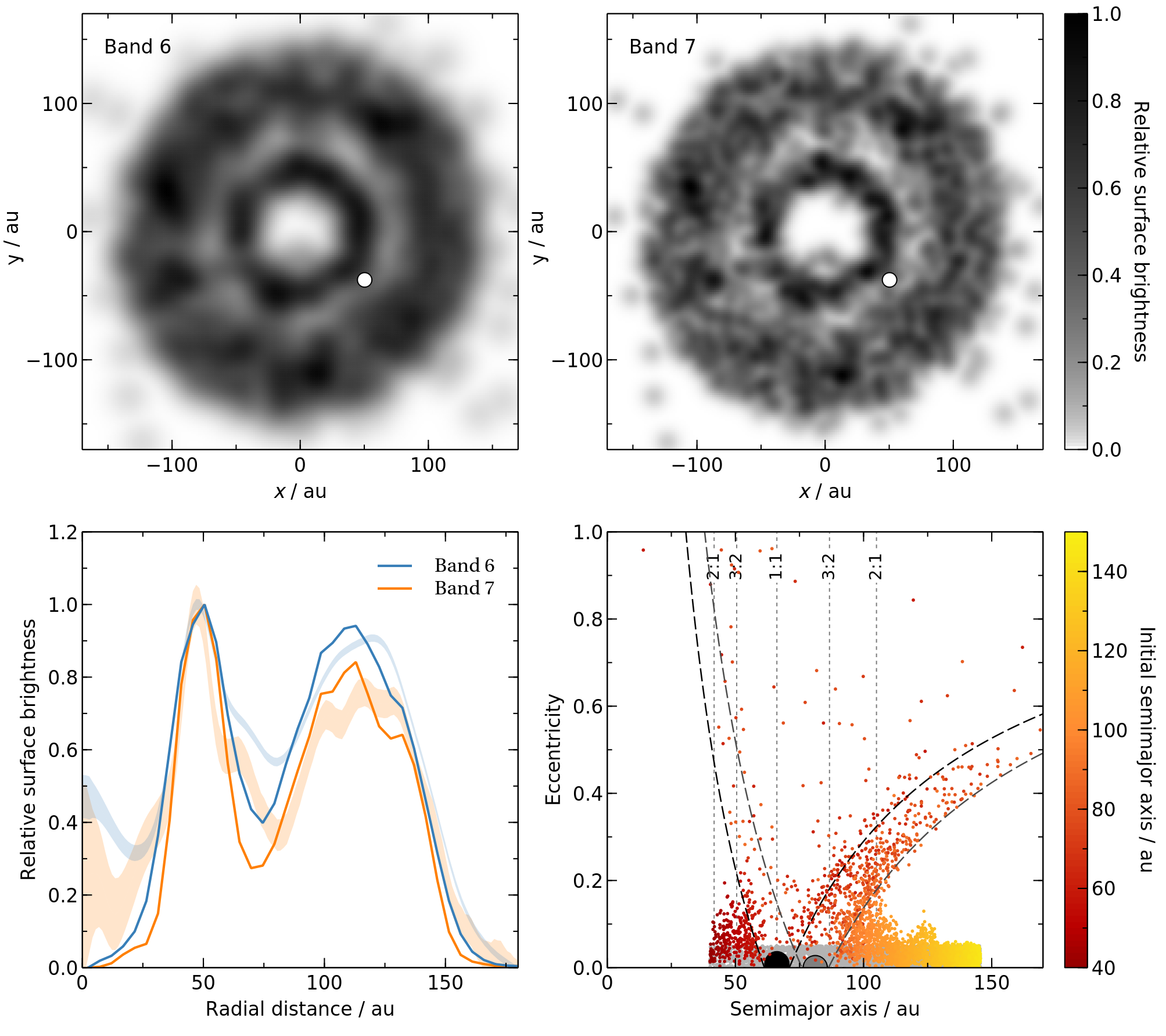}
    \caption{Simulation of ${\HD107146}$ if the debris disc gap is carved by a single migrating planet (Section \ref{subsec: hd107146}). The simulation setup had a ${0.03 \mJup}$ (${10\mEarth}$) planet initially located at ${81\au}$, embedded in a disc with initial mass ${50\mEarth}$, span ${40-146\au}$, and surface density ${\propto r^{-1/4}}$. The figure shows the simulation at ${150\myr}$, the system lifetime, over which time the planet migrated inwards by ${15\au}$ before its migration stalled. Top plots: simulated ALMA images, where the system has been rotated to the orientation of ${\HD107146}$ on the sky, each \mbox{$n$-body} particle scaled for emission, and convolved with a 2D Gaussian representing the ALMA beam (with width ${0.7 \rm \; arcsec}$ for band 6, left, and ${0.4 \rm \; arcsec}$ for band 7, right; \citealt{Marino2018}). White circles denote the planet. Bottom-left plot: scaled surface brightness profiles. Thin lines are derived from the non-rotated simulated images, while the shaded regions show the span of the ${1\sigma}$ uncertainties from ALMA \citep{Marino2018}; blue and orange denote band 6 and 7 data respectively. Bottom-right plot: semimajor axes and eccentricities from the \mbox{$n$-body} simulation. The simulation reproduces observations reasonably well; compare the figure to Figures 1 and 2 in \citet{Marino2018}, noting that we do not include an inner disc. Our simulated gap is ${\sim30\percent}$ deeper than that observed; however, computational costs prevented us from performing an exhaustive parameter search, so it is likely that better-fitting setups also exist.}
    \label{fig: hd107146}
\end{figure*}

To test these predictions, we ran several \mbox{$n$-body} simulations of the ${\HD107146}$ system (by no means a thorough parameter space exploration, because the computational cost of each simulation is high). We found that Equation \ref{eq: gapWidthApprox} overpredicts the planet masses required to carve the gap, likely due to incomplete debris clearing as described in Section \ref{subsec: caveats}. Figure \ref{fig: hd107146} shows our best-fitting simulation; a ${0.03 \mJup}$ (${10\mEarth}$) planet embedded in a disc with initial mass ${50 \mEarth}$, span ${40-146\au}$, and surface density ${\propto r^{-1/4}}$. The star  has mass ${1\mSun}$, and the planet was initially located at ${81\au}$. The simulation ran for ${150\myr}$, the age of ${\HD107146}$, during which time the planet migrated inwards by ${15\au}$ before its migration stalled. The top plots of Figure \ref{fig: hd107146} show simulated ALMA images, constructed using a similar method to \cite{Pearce2015}; we rotated the simulation to the orientation of ${\HD107146}$ on the sky, scaled the \mbox{$n$-body} particles for emission (${\propto r^{-1/2}}$), and convolved the result with a 2D Gaussian representing the ALMA beam (${0.7 \rm \; arcsec}$ for band 6 and ${0.4 \rm \; arcsec}$ for band 7; \citealt{Marino2018}). The bottom-left plot shows the azimuthally-averaged surface brightness profiles of the (non-rotated) simulation images, and the bottom-right plot the simulated particles. 

The above setup reproduces the ${\HD107146}$ disc reasonably well. The simulated widths and relative surface brightnesses of the inner and outer debris rings show good agreement with observations, and whilst our simulated gap is ${\sim30\percent}$ deeper than the observed gap, it is still reasonably consistent. In this particular simulation the material in the gap comprises mainly high-eccentricity scattered particles, many of which are detached from the planet. Few low-eccentricity Trojans are present in this simulation, but we note that stronger Trojan populations are present in some of our other ${\HD107146}$ simulations; therefore, future detections (or non-detections) of Trojans in the ${\HD107146}$ gap cannot alone confirm or deny the migrating planet scenario (for example, strong Trojan populations are visible in Figure \ref{fig: resonanceSweeping}, which shows a migrating planet). We also note that, since a small fraction of particles occupy various other resonance in this simulation, some tentative resonant structures may be visible in the simulated images. Again, whilst such features are hard to discern in this particular simulation, we note that resonant structures are much more pronounced in some of our other simulations (e.g. Figure \ref{fig: resonanceSweeping}). If resonant features were detected in the ${\HD107146}$ disc then this could support a migrating planet scenario, because debris swept into resonance by a migrating planet can form structures that are observable with current instrumentation (Booth et al., in prep.).

The simulation in Figure \ref{fig: hd107146} shows a low-mass, significantly-migrating planet, but the same gap width could also be carved by a high-mass, barely migrating planet (Figure \ref{fig: gapWidths}); a similar setup but with a more-massive (${1.5 \mJup}$) planet starting at ${76\au}$ produces a gap with very similar width {to that in Figure \ref{fig: hd107146}}, but with a deeper surface density depletion that is a worse match to observations. This is because the more-massive planet effectively removes all non-resonant debris from the gap, whilst the lower-mass, significantly-migrating planet leaves a detached population with orbits that still enter the gap, resulting in a shallower depletion (see Section \ref{subsec: insightDiscussion}). For this reason, if the gap of ${\HD107146}$ is carved by a single planet, then we favour a scenario where that planet has migrated.

Since each simulation comprises a large number of particles, and we were limited in our ability to post-process simulations (because changing the masses or initial distributions of planetesimals would fundamentally change the interaction, necessitating separate simulations for each setup considered), computational costs precluded us from exploring a broad parameter space for ${\HD107146}$. Therefore, whilst our best simulation can reproduce the observed disc reasonably well, we believe it likely that some untested migration setup would better match observations. A more thorough parameter-space search is therefore recommended for this system.

\subsubsection{${\HD15115}$, ${\HD92945}$, ${\HD206893}$ and the Solar System}
\label{subsec: otherSystems}

\begin{figure*}
    \centering
    \includegraphics[width=17cm]{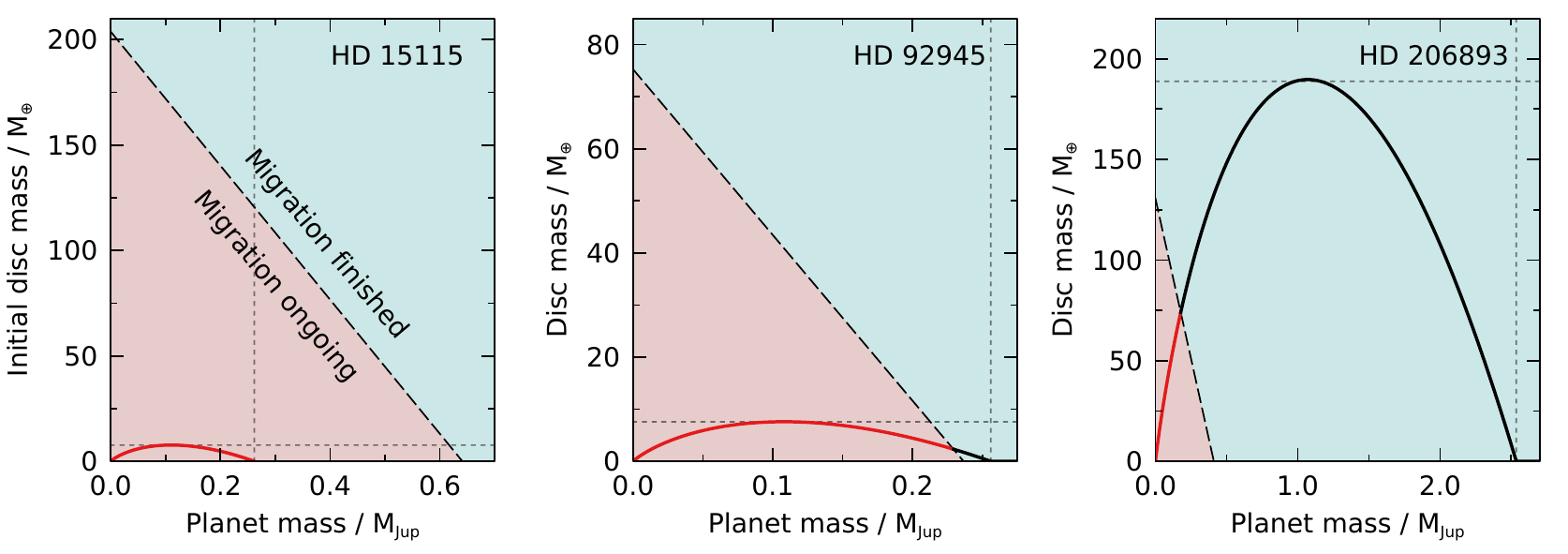}
    \caption{Constraints on the masses of planets and initial (pre-interaction) debris discs for three extrasolar systems, assuming that their gaps were each carved by an \textit{in situ} embedded planet (Section \ref{subsec: otherSystems}). For each system, the planet and initial disc masses must either lie in the red region (in which case the interaction is ongoing) or along the black section of the solid line (in which case planet migration has finished). The dotted horizontal and vertical lines respectively show the maximum initial disc masses (Equation \ref{eq: maxDiscMassForGap}) and maximum planet masses (Equation \ref{eq: gapWidthApprox} with ${M_{\rm disc}=0}$) that could result in the observed gaps, \textit{if the interactions have already finished}.}
    \label{fig: gapConstraintsForExampleSystems}
\end{figure*}

We now apply our analytic results to three other gapped extrasolar debris discs: ${\HD15115}$, ${\HD92945}$ and ${\HD206893}$. We assume that each gap was carved by an \textit{in situ} embedded planet, and we use several arguments to find upper mass limits for both the initial (pre-gap) debris discs and the responsible planets. Our resulting constraints for each system are shown in Figure \ref{fig: gapConstraintsForExampleSystems}, and we describe our methodology here. We start with ${\HD15115}$, an F-type star with an age of ${20-45\myr}$, which hosts a debris disc spanning ${44-92\au}$ with a ${14\au}$-wide gap at ${59\au}$ \citep{MacGregor2019}. First, we consider timescales; if the planet mass and/or initial disc mass are sufficiently high then the interaction would already have finished, but if they were low enough then the interaction would still be ongoing. Rearranging Equation \ref{eq: interactionTimescale}, we see that if the \textit{combined} mass of the planet and initial disc is less than ${200\mEarth}$ (${0.64\mJup}$) then planet migration would still be ongoing after ${20\myr}$ (shown by the dashed lines dividing the red and blue areas in Figure \ref{fig: gapConstraintsForExampleSystems}). Next, we consider which combinations of planet mass and initial disc mass could reproduce the observed gap, using Equation \ref{eq: gapWidthApprox}; care is required here, because the gap width predicted by this equation refers to the width at the \textit{end} of the interaction, once migration has stalled. The solid lines in Figure \ref{fig: gapConstraintsForExampleSystems} show Equation \ref{eq: gapWidthApprox} rearranged for ${M_{\rm disc}}$, assuming ${k=3}$ and ${\gamma=1.5}$. If the interaction had finished, then this equation would be very constraining; it suggests that the initial disc mass could not have been greater than just ${7\mEarth}$, because if it were larger then any planet small enough to carve such a narrow gap according to e.g. \cite{Wisdom1980} would eventually migrate and carve a wider gap before its migration stalled. Equation \ref{eq: gapWidthApprox} also implies that, if the interaction has finished, the maximum planet mass that could have carved the gap is ${0.26\mJup}$. However, the disc and planet masses predicted by Equation \ref{eq: gapWidthApprox} are actually invalid for ${\HD15115}$, because these masses are so low that the corresponding interaction timescales would exceed the stellar age (on the left plot of Figure \ref{fig: gapConstraintsForExampleSystems}, the solid line does not enter the blue region). This means that \textit{no} combination of planet mass and initial disc mass could both reproduce the gap \textit{and} mean that the interaction finishes within the stellar age. Our conclusions for ${\HD15115}$ are that, if the gap was carved by a single planet embedded in a sea of planetesimals, then the initial disc mass must have been ${\lesssim200\mEarth}$, the planet mass must be ${\lesssim0.64\mJup}$, the combined mass of the planet and initial disc is below ${200\mEarth}$ (${0.64\mJup}$), and the interaction is still ongoing (meaning the gap will broaden in the future as the planet migrates or clears a wider region). 

We now apply the same methodology to ${\HD92945}$, a K-type star with an age of ${100-300\myr}$, which hosts a debris disc spanning ${52-121\au}$ with a ${20\au}$-wide gap at ${73\au}$ \citep{Marino2019}. Our analysis is shown on the middle plot of Figure \ref{fig: gapConstraintsForExampleSystems}. For this system, the interaction would have already finished (assuming a ${100\myr}$ age) if the combined planet mass and initial disc mass exceeds ${75\mEarth}$ (${0.24\mJup}$). Unlike ${\HD15115}$, it is possible for the interaction to have finished and the observed gap to have formed, but only in a very narrow region of parameter space (if the planet mass is ${\sim 0.25\mJup}$ and the initial disc mass ${\lesssim3\mEarth}$). So for ${\HD92945}$, we can say that if the gap was carved by an \textit{in situ} embedded planet, then the initial disc mass must have been ${\lesssim75\mEarth}$, the planet mass ${\lesssim0.26\mJup}$, and the gap will continue to grow in future unless the planet mass is ${\sim 0.25\mJup}$ and the initial disc mass was ${\lesssim3\mEarth}$.

For ${\HD206893}$ it is more likely that planet migration has already stalled; this is an F-type star with uncertain age (${50\myr}$ to ${\sim 1\gyr}$), a disc spanning from ${<50}$ to ${\sim200\au}$, a ${30\au}$-wide gap centred on ${60-70\au}$, and a brown dwarf at ${10\au}$ \citep{Marino2020, Nederlander2021}. Here, the interaction would already have finished within ${50\myr}$ if the combined planet mass and initial disc mass were greater than ${130\mEarth}$ (${0.41\mJup}$). The right plot of Figure \ref{fig: gapConstraintsForExampleSystems} shows that many combinations of planet mass and initial disc mass would allow the interaction to have finished and the observed gap to have been carved; for ${\HD206893}$ we can therefore say that, if the gap was carved by an \textit{in situ} embedded planet, then the initial disc mass must have been ${\lesssim190\mEarth}$, the planet mass ${\lesssim2.5\mJup}$, and planet migration has now stalled unless the combined planet mass and initial disc mass were less than ${130\mEarth}$ (${0.41\mJup}$).

For ${\HD15115}$ and ${\HD92945}$, the maximum disc masses inferred from our gap-clearing arguments (${200\mEarth}$ and ${75\mEarth}$ respectively) are smaller than the lower limits required for self-stirring (${290\pm30\mEarth}$ and ${290^{+80}_{-90}\mEarth}$ respectively; \citealt{Pearce2022}). This could imply that these discs are stirred by some other mechanism, e.g. by planet(s) in the gaps (for ${\HD206893}$, the same method finds that the disc could self-stir within ${1\gyr}$ if its mass were ${\gtrsim120 \mEarth}$, which is consistent with our gap-clearing limit of ${\lesssim190\mEarth}$). If we believe that these discs are actually more massive than our gap-clearing estimates, then there are a number of possibilities; either these gaps were not carved by individual embedded planets, or much of the material interior to the gaps is in resonance with the migrated planet (Section \ref{subsec: resonances}), or the gap locations were already at least partially depleted of material when planet formation ceased (Section \ref{subsec: caveats}).

Finally, we apply our analysis to the Solar System; specifically, we consider whether dwarf planets embedded in the Solar System's debris discs (Ceres in the Asteroid Belt, and Pluto in the Kuiper Belt) could significantly migrate by scattering debris. Ceres, with mass ${1.6 \times 10^{-4}\mEarth}$ and semimajor axis ${2.8\au}$, is embedded in the Asteroid Belt of total mass ${\sim4\times 10^{-4}\mEarth}$ \citep{Pitjeva2018} spanning roughly ${2.1-3.3\au}$. Equation \ref{eq: discMassForMigration} shows that Ceres should undergo significant migration if its mass is less than 5 times the mass of the remaining belt, as indeed is the case. However, it is likely that the migration timescale would be so long that it renders this effect negligible; Equation \ref{eq: interactionTimescale} shows that such migration would take ${\sim10^{16}\yr}$ (much longer than the Solar System lifetime), so whilst Ceres would migrate given infinite time, for practical purposes it does not migrate due to this effect. Pluto is more complicated, since its orbital evolution is dominated by its resonance with Neptune, but for this simple example we will neglect this. Pluto has a mass of ${2\times 10^{-3} \mEarth}$ and is located at the inner edge of the classical Kuiper Belt, where the latter has total mass ${\sim0.01\mEarth}$ \citep{Fraser2014} roughly located between the 3:2 and 2:1 resonances with Neptune (${40-48\au}$). Equation \ref{eq: discMassForMigration} shows that Pluto would undergo significant migration if its mass were less than 10 times the mass of the remaining belt, as is again the case. However, once again the migration rate would be negligible; Equation \ref{eq: interactionTimescale} shows that migration would take ${\sim10^{15}\yr}$. In summary, the present-day masses of the Solar System's debris discs are too low to cause significant dwarf-planet migration\footnote{The early Kuiper Belt may have been more massive than today, before Neptune scattered much of its material \citep{Gomes2005}. This occurred around the Late Heavy Bombardment (LHB), ${\sim 700 \myr}$ after planet formation \citep{Hartmann2000}. Equation \ref{eq: interactionTimescale} shows that, for Pluto to have undergone significant planetesimal-driven migration before the LHB, the early mass of the Kuiper Belt would need to have been $\gtrsim 20\mEarth$, i.e. ${\gtrsim 2000}$ times its present-day mass. This is consistent with Kuiper-Belt origin hypotheses \citep{Morbidelli2020}, so it is possible that planetesimal-driven migration drove Pluto inwards before the LHB.}.

\subsection{\textit{Caveats}}
\label{subsec: caveats}

There are a number of \textit{caveats} that the reader should be aware of before applying our results. Firstly, our analyses only apply if the large body (`planet') is much more massive than each \textit{individual} planetesimal (although the \textit{total} mass of planetesimals can exceed that of the large body). This is because we model the interaction as a large body moving through a `sea' of much smaller objects, so the interaction acts like a smooth drag force on the large body. If the migrating body were instead embedded in a disc of similar-sized bodies, then the evolution would be much more stochastic in nature (each individual encounter would be more akin to a planet-planet scattering event), and the results could differ from ours. For this reason, our analysis applies to planets and dwarf planets embedded in debris discs, but not to individual planetesimals.

Secondly, we assumed that the interaction starts with a planet embedded in a smooth planetesimal disc with a constant surface-density index (Figure \ref{fig: cartoon}). However, if the planet formed \textit{in situ} then the initial surface density may be reduced around the planet, because some solids located near the planet in the protoplanetary disc would have accreted onto the planet during its formation. Gaps observed in protoplanetary discs could signify where disc material has already been incorporated into forming planets \citep{Andrews2018, Lodato2019}, although this picture is complicated and alternative, non-planetary explanations also exist for these gaps (e.g. \citealt{Okuzumi2016, Dullemond2018}). The accretion of material during planet formation could allow small planets to inhabit narrower gaps in a debris disc than our models suggest, because the material depletion surrounding the planet would weaken its initial, planetesimal-driven migration. However, if a small planet can somehow begin planetesimal-driven migration, then it would encounter `fresh' material and its migration could be sustainable, and our results would hold. This could occur if the region surrounding the planet is not significantly depleted during planet formation, or if planetesimals move into that region after formation (for example, material that has been scattered by other planets), or if the planet moves into a debris disc during a period of dynamical upheaval (such as Neptune entering the Kuiper Belt in the early Solar System: \citealt{Gomes2005, Tsiganis2005}).

Thirdly, whilst our models include the back-reaction of planetesimals on the planet, they omit disc self-gravity (i.e. the planetesimals do not interact with each other). This could be significant if the planet is much less massive than the disc, where the force on a planetesimal resulting from the disc may be greater than that from the planet. The effects of omitting self-gravity are unclear, and require more detailed models to confirm; it could be, for example, that self-gravitating material could resist being swept into resonances, or that scattered particles which would otherwise be ejected would in reality be damped and retained.

Another \textit{caveat} is that we assumed the planet opens up a gap as it migrates. However, this might not be the case for very low-mass planets; if the debris-ejection timescale is longer than the migration timescale, then the planet would migrate inwards faster than it could clear a gap. In this regime only a shallow gap would be formed, and our approximation that the planet ejects all debris it encounters would not be correct (see below). We therefore urge caution when applying our equations to rapidly migrating planets, since in this case the gaps may not be as deep or as broad as we predict.

Additionally, our equations involving the predicted gap width (Equations \ref{eq: gapWidthBothNotation}-\ref{eq: maxDiscMassForGap}) assume that the interaction has already finished; they describe the \textit{post-interaction} gap widths, once the planet has stopped migrating. For an observed gapped disc, it is possible that an embedded planet is actually still migrating, in which case the gap will widen in future and Equations \ref{eq: gapWidthBothNotation}-\ref{eq: maxDiscMassForGap} may not be valid. We describe a way to check and account for this in Section \ref{subsec: otherSystems}.

Finally, our predicted gap widths can be overestimates in some regimes, for three reasons. One reason is that our analytic models assume that the planet eventually ejects all material that comes close to it. In reality, whilst the planet would scatter nearby material, it may migrate and decouple from scattered debris before those particles have undergone sufficient interactions to eject them. This leaves a detached debris population, such as those visible in Figures \ref{fig: simulationSetup}, \ref{fig: resonanceSweeping} and \ref{fig: hd107146} (the non-resonant particles lying between the grey and black dashed lines on the semimajor axis-eccentricity plots). Since the planet fails to eject these particles, it does not lose as much energy as our analytics assume, and so it does not migrate as far. This means that the gap widths that we predict can be overestimates if the planet migrates fast enough to decouple from scattered debris. Another reason is that our analytic gap widths refer to gaps in \textit{semimajor axis} space, rather than radial space (since they are predicted from energy arguments, and energy is set by semimajor axis). Since debris particles have non-zero eccentricities (either low eccentricities in the cold population, or high eccentricities in the scattered population), this has the effect of radially smoothing the gap edges, and so our predicted gap widths will be wider than what would be measured from an observed radial surface-brightness profile. Finally, resonance sweeping can also trap particles, which sometimes results in narrower gaps than we predict (Section \ref{subsec: resonances}). For these reasons, in some cases the gaps we predict may be wider than what would actually be observed.

\subsection{Potential insights into debris disc masses, planet properties and system evolution}
\label{subsec: insightDiscussion}

Our results could yield insights into the properties and evolution histories of debris discs and planetary systems. Firstly, a major problem in the field is that we do not know the masses of debris discs, which is arguably their most fundamental attribute. Whilst estimates are found by extrapolating the mass in visible dust up to the most-massive bodies, this process is highly uncertain owing to a lack of constraints on the size distribution and size of the largest objects, and small changes in assumptions can cause estimates to vary by orders of magnitude. This process can produce results that are incompatible with our ideas of disc formation (the `debris disc mass problem'; see \citealt{Krivov2021}), which in turn raises significant questions about the process by which debris evolved and gets `stirred' such that observable dust is released through violent collisions between larger bodies (\citealt{Pearce2022}; see also \citealt{Najita2022}). Accurate knowledge of debris-disc masses is therefore imperative for understanding these objects, but there are too many unknowns in how observable dust relates to the total debris population for firm conclusions to be drawn from this technique alone.

A way to break this degeneracy is to determine disc masses indirectly, by considering processes that depend on the disc mass but not on the unknown size distribution or maximum-size cutoff. A clear example is an interaction between a planet and a debris disc, where the back-reaction of the disc mass causes the planet to evolve too. It is already known that a massive debris disc would drive planetary precession, which in turn induces disc structures that would be difficult to explain via other processes \citep{Pearce2015, Sefilian2021}; in such interactions the planet precession rate (and thus the location of these structures) is set by the disc mass, so observations of such structures (or non-detections, if coupled with a known planet) could strongly constrain disc mass. That was a secular effect, but a similar argument applies to the scattering interaction we study here; we showed that small planets embedded in massive debris discs would carve wide gaps, predicted gap width as a function of disc and planet mass, and showed that there is a maximum disc mass that can result in a given gap width that is independent of planet mass, so a detection (or non-detection) of a planet in a gap could yield valuable constraints on disc mass. In general, we argue that planet-disc interactions are a promising tool for providing independent estimates of debris disc masses.

With our results, one could also use observations of a gapped debris disc to infer the properties and evolutionary history of an unseen planet. We showed that the same gap could often be carved by two different planets: either a higher-mass, barely-migrating planet, or a lower-mass, strongly-migrating planet. The planet masses in each case could be constrained using our equations (combined with assumptions about the disc mass). However, whilst the gap width is identical in the two scenarios, the debris parameters are not; Figure \ref{fig: simulationSetup} shows that for a low-mass, strongly-migrating planet (bottom plot), surviving debris is much more excited than if the gap were carved by a high-mass, barely-migrating planet (top plot), and there is a stronger scattered component in the former case. The higher level of debris excitation in the first case results from mean-motion resonances sweeping through the disc as the lower-mass planet migrates (Section \ref{subsec: resonances}), which pumps particle eccentricities and could potentially result in increased debris collisions and dust production. The stronger scattered disc arises not only because the lower-mass planet is less efficient at ejecting debris than the higher-mass one, but also because as it migrates it decouples from scattered material, leaving behind a detached population of high-eccentricity debris (which would smooth the observed disc outer edge, e.g. \citealt{Faramaz2021}). So if a gapped disc were found to have either a high debris-excitation level (which could be observable, e.g. \citealt{Marino2021}), or strong evidence of scattered disc, then this would point towards a lower-mass, significantly migrating planet as the gap carver. A shallow gap in an older disc could also be evidence of a migrating planet, since in this case detached material would still be present in the gap at pericentre, whilst a massive, stationary planet would eventually clear all material and leave a deep gap. The gap edges may also be smoother if carved by an inward-migrating planet, and sharper if carved by a non-migrating planet, due to higher debris eccentricities in the former case (although the inner edge could be complicated by material in resonance; Figure \ref{fig: resonanceSweeping}). Finally, if a planet were actually detected, then its location could allow us to differentiate between migrating and non-migrating scenarios; a non-migrating planet would be located at the gap centre, whilst a migrating planet would be located closer to one edge of the gap (Figure \ref{fig: simulationSetup}). In summary, whilst gap widths may be similar if carved by migrating and non-migrating planets, there are several potential ways to distinguish between these scenarios.

Finally, our results allow us to consider one last scenario, where a planet embedded in a debris disc migrates all the way to the disc inner edge. This can occur if the disc is sufficiently massive, where the exact mass required depends on the planet's mass and initial location, and the initial properties of the debris disc (Equation \ref{eq: migrationToInnerEdge}). In this scenario, if the planet efficiently scatters material as it migrates inwards, then the eventual outcome would be a non-gapped debris disc (with its inner edge near the initial location of the planet), and a planet residing far interior to the current disc. This means that discs which are narrower and gapless today could have been broader and gapped in the past, if an initially embedded planet has since removed all but the outermost regions of the original disc. It also means that, even if a planet has sculpted the inner edge of a debris disc, that planet does not necessarily reside close to the disc today; the responsible planet could since have migrated much further inwards due to planetesimal scattering. Alternatively, this configuration could also occur if young, massive planets rapidly sculpt nearby debris before migrating inwards during the protoplanetary disc phase (as discussed by \citealt{Pearce2022}). The \textit{James Webb Space Telescope} should soon start searching for planets at the inner edges of debris discs; if planets are not found at the expected locations then it does not necessarily mean that debris discs are not sculpted by planets, because the responsible planets could now reside much further inwards.

\section{Conclusions}
\label{sec: conclusions}

We investigated the gaps that would be carved by a single planet embedded in a massive debris disc. In this case, the planet would migrate while scattering debris particles. We provided a simple criterion for the planet-to-disc mass ratio at which planet migration becomes significant, and gave formulae for the size of the gap that a planet would carve in a massive debris disc. We also identified the interaction timescales. We found a degeneracy that is important when considering debris-disc gaps carved by embedded planets; the same width gap could be carved by either a high-mass, barely-migrating planet, or a low-mass, significantly-migrating planet. We showed that, counterintuitively, a low-mass planet could carve a wider gap than a high-mass planet, and that there is a minimum gap width that a single planet can carve. Our analytic predictions were verified using \mbox{\textit{n}-body} simulations. We also found that resonance sweeping can be very important in this scenario, affecting the planet migration distance, disc structure, and debris excitation level. In particular, we showed that the inner edges of gapped debris discs could actually consist of debris swept into mean-motion resonances by an inwardly-migrating planet. We applied our results to various systems, and showed that the disc of ${\HD107146}$ can be reasonably reproduced if a migrating planet inhabits the gap. Our results  show that debris-disc masses should not be ignored when considering gaps carved by embedded planets, and that constraints from planet-disc interactions could yield significant insights into debris-disc masses and system evolution.

\section*{Acknowledgements}

\noindent We thank Sebasti\'{a}n Marino for sharing his ${\HD107146}$ surface brightness profiles, and the anonymous referee whose detailed review significantly improved the paper. TDP and TL are supported by the Deutsche Forschungsgemeinschaft (DFG) grants ${\rm Kr \; 2164/14{\text -}2}$, ${\rm Kr \; 2164/15{\text -}2}$ and ${\rm Lo \; 1715/2{\text -}2}$.

\section*{Data Availability}

The data underlying this article will be shared upon reasonable request to the corresponding author.



\bibliographystyle{mnras}
\bibliography{bib_friebeGapsInDebrisDiscs}





\bsp	
\label{lastpage}
\end{document}